\documentclass{elsart}
\usepackage[usenames]{color}  %PTW
\usepackage{ifthen}  %PTW
\usepackage{natbib}
\usepackage{epsfig}

\usepackage{amssymb}
\newcommand{\beq}{\begin{equation}}
\newcommand{\eeq}{\end{equation}}

\newcommand{\bea}{\begin{eqnarray}}
\newcommand{\eea}{\end{eqnarray}}

\newcommand{\beal}{\begin{mathletters}\begin{eqnarray}}
\newcommand{\eeal}{\end{eqnarray}\end{mathletters}}

\newcommand{\p}{\partial}
\newcommand{\del}{\nabla}
\newcommand{\grad}{\nabla}

\newcommand{\ucd}{{D}^{[u]}_t}
\newcommand{\ucid}{\ucd}

\newcommand{\sR}{s_{{}_R}}
\newcommand{\sM}{s_{{}_M}}
\newcommand{\aR}{a_{{}_R}}
\newcommand{\aM}{a_{{}_M}}
\newcommand{\bRM}{b_{{}_{RM}}}
\newcommand{\bMR}{b_{{}_{MR}}}

\newcommand{\am}{A^{(-)}}
\newcommand{\ap}{A^{(+)}}

\newcommand{\Rey}{\mathbf{R}}
\newcommand{\rey}{\Rey}
\newcommand{\Max}{\mathbf{M}}
\newcommand{\Lop}{\mathbb{L}}

\ifthenelse{1=1}  % Here are some shortcuts to creating color.
{

\newcommand{\green}{\color{OliveGreen}}
\newcommand{\grey}{\color{Gray}}
} {

\newcommand{\green}{}
\newcommand{\grey}{}
}

\begin{document}
\begin{frontmatter}


\title{Turbulent Magnetohydrodynamic Elasticity: \\
I. Boussinesq-like Approximations for Steady Shear} % \thanksref{label1}}
\author{Peter Todd Williams} % \corauthref{cor1}\thanksref{label2}}
\ead{ptw@lanl.gov}
\address{Los Alamos National Laboratory, MS P225, P.O. Box 1663, Los Alamos, NM 87545, USA}%\thanksref{label3}}

%%%%%%%%%%%%%%%%%%%%%%%%%%%%%%%%%%%%%%%%%%%%%%%%%%%%%%%%%%%%%%%%%%%%%%%%%%%%%%%%
\begin{abstract}
We re-examine the Boussinesq hypothesis of an effective turbulent viscosity within
the context of simple closure considerations for models of strong magnetohydrodynamic turbulence.
Reynolds-stress and turbulent Maxwell-stress closure models will necessarily introduce a suite
of transport coefficients, all of which are to some degree model-dependent. One of the
most important coefficients is the relaxation time for the turbulent Maxwell stress,
which until recently has been relatively ignored. We discuss this relaxation within the
context of magnetohydrodynamic turbulence in steady 
high Reynolds-number, high magnetic-Reynolds-number shearing flows. 
 The relaxation time for the turbulent Maxwell stress
is not limited by the shear time scale, in contrast with Reynolds-stress closure
models for purely hydrodynamical turbulence. The anisotropy of the turbulent stress tensor
for magnetohydrodynamic turbulence, even for the case of zero mean-field considered here,
can therefore not be neglected. This shear-generated anisotropy can be interpreted as being
due to an effective turbulent elasticity, in analogy to the Boussinesq turbulent viscosity. 
We claim that this turbulent elasticity should be important for any astrophysical problem in
which the turbulent stress in quasi-steady shear has been treated phenomenologically with an
effective viscosity.
\end{abstract}
%%%%%%%%%%%%%%%%%%%%%%%%%%%%%%%%%%%%%%%%%%%%%%%%%%%%%%%%%%%%%%%%%%%%%%%%%%%%%%%%

\begin{keyword}
\end{keyword}

\end{frontmatter}

%%%%%%%%%%%%%%%%%%%%%%%%%%%%%%%%%%%%%%%%%%%%%%%%%%%%%%%%%%%%%%%%%%%%%%%%%%%%%%%%
% main text
%%%%%%%%%%%%%%%%%%%%%%%%%%%%%%%%%%%%%%%%%%%%%%%%%%%%%%%%%%%%%%%%%%%%%%%%%%%%%%%%

\section{Introduction}

The Boussinesq hypothesis of an effective turbulent
viscosity is still widely used both in engineering applications and in astrophysics. 
This is true despite the passage of over one hundred years since the birth of the concept.
There are a very large number of 
models for turbulent stress that give substantially better results than a simple effective viscosity, and in some situations, such
as strongly time-dependent flows, the use of an effective viscosity gives notoriously bad results. Despite this, the
concept of an effective turbulent viscosity has proven to be so persistent because it is useful, simple, and because it is based on an
appealing analogy between the random motions of turbulent blobs and the kinetic theory of thermal motions of molecules.
Thus, effective turbulent viscosity is commonly invoked as a good
crude ``lowest-order'' approximation, for problems in which the Reynolds number is very large.

For problems in which the magnetic Reynolds number is also very large (as is often the case in astrophysics), we ask if a picture of 
turbulence as colliding blobs of fluid is at all useful even as a crude approximation. A more appropriate analogy
for the turbulent component of the {\em field}  may be not the kinetic theory of point-like molecules, but rather
the kinetic theory of long polymers in melt or in solution \citep{Ogil:2001, Will:2001}. 
For the crudest possible model, such an analogy would
suggest the concept of an effective turbulent elasticity, in addition to the effective turbulent viscosity. Thus,
corresponding to the Boussinesq approximation in hydrodynamic turbulence, the corresponding ``lowest-order'' approximation
for the stress in MHD turbulence should include an additional coefficient, in analogy to a simple viscoelastic
fluid. Just as the inclusion of elasticity in the stress response of an ordinary laboratory fluid can have dramatic
consequences, we suggest that {\em any} astrophysical problem in which turbulence has been treated phenomenologically
with the Boussinesq viscosity might have dramatically different solutions if an effective elasticity were included as well.

%Hydrodynamical turbulence and magnetohydrodynamical turbulence are quite different. 
In particular, for the purposes of quasi-steady shear, we claim
%one of the salient differences between the two is 
that the anisotropy of the stress tensor 
%in response to shear 
is of much greater significance in the case of MHD turbulence than in the case of purely hydrodynamic turbulence.
Thus, while investigations of the idealized problem of homogeneous istotropic
hydrodynamic turbulence may lead to a greater understanding of the dynamical importance of turbulence
to the large-scale mean flow of a turbulent fluid, we expect that, quite broadly speaking, models
for MHD turbulence that are to be of some use in predicting the behavior of the mean flow must 
address directly the anisotropy of the turbulent Maxwell stresses, irrespective of the presence
or absence of any mean field.

Many of the more recent simulations of MHD turbulence in a shearing envoronment (our main interest in this paper)
are simulations (e.g. \citet{HaGaBa:1995}) of the Velikov-Chandrasekhar-Balbus-Hawley magnetorotational instability (MRI), and we will
refer to the results of these simulations for guidance. See also \citet{Will:2002} for a comparison of
simulations of the MRI with some viscoelastic models.

For our purposes, the dominant feature of simulations of the MRI is that the largest component of the turbulent
stress tensors is {\em not} the ``viscous'' $r\theta$ component, but rather the streamwise $\theta\theta$ component. 
(Note that in simulations of the MRI, $\theta$ is the direction of the shear, and $r$ is the cross-shear direction.)
%The first has
%a ready interpretation as an effective viscosity; the latter may be interpeted as an effective elasticity.
In fact, our original reason for investigating viscoelastic models was as a mechanism to collimate and
possibly drive jets using this
$\theta\theta$ component of the stress tensor.

Most importantly we wish to point out that the interpretation of the turbulent stress as an effective
viscous stress can lead to qualitatively erroneous predictions. In view of the fact pointed out above
that the ``viscous'' component is in fact not even the largest component of the stress tensor, a
prescription for the turbulent stress tensor $\Pi^{turb}$ of the form
\beq
\Pi^{turb}_{ij} = \mu_{\rm turb} \left[\p_i v_j + \p_j v_i - {2 \over 3} (\p_k v_k) \delta_{ij} \right] 
+ \mu^{[b]}_{\rm turb} (\p_k v_k) \delta_{ij},
\eeq
where $\mu_{\rm turb}$ and $\mu^{[b]}_{\rm turb}$ are effective shear and bulk viscosities respectively,
is untenable, even as a gross approximation.

One possible approach to the problem is to generalize the viscosity to a tensor viscosity. This
approach ignores the fact, however, that the large azimuthal component of stress in MRI-driven
shear turbulence is fundamentally due to the vector advection properties of the induction equation.
%We suspect that this ``elastic'' stress is
%in fact a fairly reliable feature of any MHD turbulence against a background of shear.

\section{Notation}
As is conventionally done in the Reynolds decomposition in hydrodynamic turbulence, it is convenient here to
decompose all fluid variables into mean and fluctuating, or ``turbulent,'' parts. The former will be denoted with 
an overbar, and
the latter with a prime. Thus, for example, $B_i = \bar B_i + B'_i$. 
We employ standard ensemble rules of averaging, so that averaging commutes with time and spatial derivatives.
It is also useful to think of this averaging as, say, a time or spatial averaging over some convenient timescale or lengthscale,
although strictly speaking this is incorrect, because such averaging introduces additional terms.

Furthermore, for the sake of definiteness,
we will make liberal use of Cartesian index notation for vectors and tensors, although we may occasionally
use abstract vector notation when there is no chance of ambiguity.
To make lengthy equations a bit more palatable, we denote symmetrization and antisymmetrization on indices
with curly and square brackets respectively, so that, e.g.,
\[
A_{\{ij\}} \equiv A_{ij} + A_{ji}.
\]
The generalization of the given
equations to curvilinear (``covariant and contravariant'') form is straightforward.

We use
Heaviside-Lorentz units for the magnetic field. 
The use of Gaussian units and the attendant factors of $1/\sqrt{4 \pi}$ 
 would make lengthy equations even longer, and this would detract from readability.
The conversion to Gaussian units is
\[
   B_i^{[Gauss]} = \sqrt{4 \pi} B_i^{[Heaviside-Lorentz]}.
\]

The full Maxwell stress tensor, under the approximation that $|E| \ll |B|$, has
components  $B_iB_j - \half B^2 \delta_{ij}$.
The turbulent Maxwell stress is the Maxwell stress due to the turbulent component of the field, $B'$.
We find it more convenient to work with the magnetic cross-correlation tensor $M_{ij}$ where
$M_{ij} \equiv \overline{ B'_i B'_j}$. Then the turbulent Maxwell stress is
\[
\Max_{ij} \equiv M_{ij} - M_{kk} \delta_{ij}.
\]

\section{Invariant Tensor Advection Operators}
In perfect flux-freezing the stress tensor $M_{ij} \equiv B'_iB'_j$ evolves by
the action of the (modified) upper-convected invariant derivatve $\ucd$ such that $\ucd(M)_{ij} = 0$,
where
\beq
(\ucid M)_{ij} 
\equiv
(\partial_t + \bar v_k \partial_k)M_{ij} - (\partial_k \bar v_i)M_{kj} - 
M_{ik}(\partial_k \bar v_j) + 2(\partial_k \bar v_k)M_{ij}.
\label{eqn:ucm}
\eeq
(The required modification of the conventional upper-convected invariant derivative 
is the inclusion of the term proportional to the divergence of the velocity).
This relationship is simply a direct result of the vector advection equation for the
magnetic field in perfect MHD, and does not rely on breaking the field into mean and fluctuating
parts, although we have done so here.
This equation shows how, for example, the stress $M_{ij}$ is distorted by shear (see figure 1).

Generally speaking, tensors formed from a dyad of vectors that obey vector
advection equations will often produce a tensor advection equation that is similar to the above:

Consider the scalar advection operator
\[
\partial_t + v_k \partial_k.
\]
This may be generalized to a vector advection operator in more than one way. Note that the
vector advection operator acting on either the field $B$ in MHD or the vorticity $\omega$ in simple
hydrodynamics is, in the incompressible case (writing the vector as $w$),
\[
\partial_t w_i + v_k \partial_k w_i - w_k \partial_k v_i,
\label{eq:magadvect}
\]
reflecting that these vectors are stretched and aligned with the shear.
In contrast, in the Reynolds decomposition, 
the advection operator acting on the turbulent velocity field $v'_i$ is
\[
\partial_t v'_i + \bar v_k \partial_k v'_i + v'_k \partial_k \bar v_i + v'_k \partial_k v'_i
\]
The last term is the origin of the famous triple-correlation; here we focus on the penultimate
term, which is similar to the last term in expression (\ref{eq:magadvect}) except for a change of sign.
Certainly more complicated examples are possible. For instance,
the advection of the gradient  $g_i \equiv \partial_i \phi$ of a passive scalar $\phi$ is
\[
\partial_t g_i + v_k \partial_k g_i + g_k \partial_k v_i - \epsilon_{ijk} g_j \omega_k = 0
\]
In all cases above, however, there is a term $w_k \partial_k v_i$ that, depending upon whether
it appears with a plus sign or a minus sign, tends to anti-align or align the vector $w_i$ with the shear.
Thus, if the stress tensor is formed from vectors that obey such equations, the stress will necessarily
be distorted by the Lagrangian deformation of the fluid. Upper- and lower-convected derivatives are two
tensor generalizations of the advection operator that arise when the tensors are formed from vectors
that are aligned or anti-aligned by the shear.

\begin{figure}[htp]
\centering
%\vspace{30mm}
\epsfxsize=4.5in % will enlarge or reduce the postscript figures based on
                % the xsize
\epsfbox{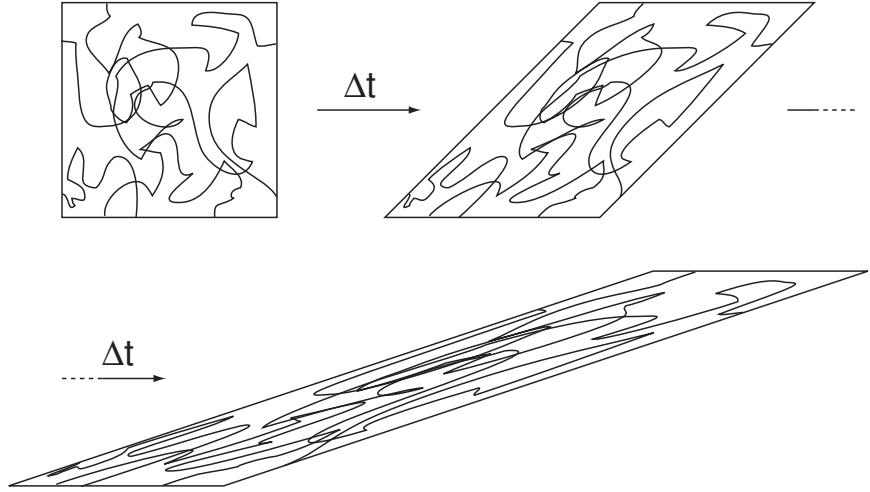}
\caption{The turbulent magnetic field lines are distorted by shear. In equilibrium, the degree of
anisotropy is determined by the balance of production, distortion, and dissipation, as described
below.}
\label{fig:fig1}
\end{figure}

One finds then that, for example, the Reynolds stress is distorted by the background shear by the
advection operator (the lower-convected invariant derivative) which, in the incompressible case,
may be written as
\beq
\partial_t \rey_{ij} + \bar v_k \partial_k \rey_{ij} + (\partial_k \bar v_i) \rey_{kj} +
\rey_{ik} (\partial_k \bar v_j).
\eeq
The second and third terms in this operator definition have long been recognized as being quite
important to the theory of hydrodynamical turbulent transport processes. These terms distort the
Reynolds stress, and are as well a source for the Reynolds stress. The distortion that shear induces
on $\rey_{ij}$ is opposite to that induced on $M_{ij}$, and in such a manner as to produce a
``viscous''-like component to the stress. 
%If the direction of the shear is $x$ and the cross-shear direction is $y$, the viscous component is the $x$--$y$ component.
 As the turbulent Maxwell stress and the Reynolds
stress appear with different signs in the turbulent stress tensor,
\beq
\Pi^{\rm turb}_{ij} = - \rey_{ij} + \Max_{ij},
\eeq
the signs of the viscous component of the Maxwell stress and the Reynolds stress are the same.
It is of course just such a distortion of the stress tensor by the deformation of the fluid (including shear)
that is the origin of {\em molecular viscosity}, where in that case the stress tensor is formed from the sums of
the contributions $mv_iv_j$ to the stress tensor from individual particles.

%%%%%%%%%%%%%%%%%%%%%%%%%%%%%%%%%%%%%%%%%%%%%%%%%%%%%%%%%%%%%%%%%%%%%%%%%%%%%%%%

\section{Reynolds Decomposition}
It is helpful to write down in full the Reynolds decomposition for MHD turbulence. Here let us concentrate
on the constant-density case, $\rho = \bar \rho$. The continuity equation and ``magnetic Gauss's law'' are then
\[ \partial_k \bar v_k = \partial_k v'_k = 0 \]
\[ \partial_k \bar B_k = \partial_k B'_k = 0 \]
The mean-field velocity transport equation is
\bea
\bar \rho \partial_t \bar v_i + \bar \rho \bar v_k \partial_k \bar v_i  = - \partial_i \bar P + 
\bar B_k \partial_k \bar B_i - \half \partial_i (\bar B^2) - \nonumber \\
\partial_k R_{ik} + \partial_k M_{ik} - \half \partial_i M_{kk} + \mu\p_{kk}\bar v_i
\nonumber
\eea
The unmodeled transport equation for the Reynolds stress is obtained after some tedious but straightforward
algebra:
\bea
\partial_t R_{ij} + \bar v_k \partial_k R_{ij} + (\partial_k \bar v_i)R_{kj} + R_{ik}(\partial_k \bar v_j)
+ \bar \rho \partial_k (\overline{ v'_i v'_j v'_k }) = 
\nonumber \\
- \overline{v'_{\{i} \partial^{}_{j\}} P'} + 
(\partial_k \bar B^{}_{\{i}) \overline{v'_{j\}} B'_k} + 
\bar B_k \overline{(\partial_k B'_{\{i} ) v'_{j\}} } +  
\overline{B'_k (\partial_k B'_{\{i}) v'_{j\}}} -
\nonumber \\
- \overline{v'_{\{i}\left(\partial^{}_{j\}}(\bar B_k B'_k)\right)} - 
\half \overline{v'_{\{i}\left( \partial^{}_{j\}}(B'_k B'_k)\right)}  
+ \mu \left(\partial^2_{k}R_{ij} - 2 \overline{(\partial_k v'_i)(\partial_k v'_j)} \right)
\label{eq:rey}
\eea
%\bea
%\partial_t R_{ij} + \bar v_k \partial_k R_{ij} + (\partial_k \bar v_i)R_{kj} + R_{ik}(\partial_k \bar v_j)
%+ \bar \rho \partial_k (\overline{ v'_i v'_j v'_k }) = 
%\nonumber \\
% - \left[ \overline{v'_i \partial_j P'} + \overline{v'_j \partial_i P'} \right]
%+ 
%{\red
%(\partial_k \bar B_i) \overline{v'_j B'_k} + \overline{v'_i B'_k} (\partial_k \bar B_j) }
%+ 
%\nonumber \\
%{\vermil
%+
%\bar B_k \overline{(\partial_k B'_i ) v'_j } + \bar B_k \overline{v'_i (\partial_k B'_j)}
%} + {\prored \overline{B'_k (\partial_k B'_i) v'_j} + \overline{B'_k v'_i (\partial_k B'_j)}
%}
%-
%\nonumber \\
%{\blue
%\overline{\left(\partial_i(\bar B_k B'_k)\right) v'_j} - \overline{v'_i\left( \partial_j(\bar B_k B'_k)\right)} }
%- 
%{\ltblue \half \overline{\left(\partial_i(B'_k B'_k)\right) v'_j} - \half \overline{v'_i\left( \partial_j(B'_k B'_k)\right)}  
%}
%\nonumber \\
%+ \mu \left(\partial^2_{k}R_{ij} - 2 \overline{(\partial_k v'_i)(\partial_k v'_j)} \right)
%\eea
There are three magnetic curvature terms, namely those arising from
$B' \cdot \grad \bar B$, those from $\bar B \cdot \grad B'$, and those from $B' \cdot \grad B'$.
Likewise, there are two magnetic pressure terms, namely those arising
from the cross-correlation pressure $\bar B B'$ and those from the pure turbulent field
pressure $B'B'$. It should be noted that in all terms in which both the velocity and the field appear,
the velocity appears once and the field appears twice.

We write the induction equation as
\[
\partial_t B_i + v_k \partial_k B_i - B_k \partial_k v_i + B_i (\partial_k v_k) = \eta \partial^2_k B_i.
\]
%and in what follows we assume that $\partial_k v_k = 0$.
From the induction equation we find that the mean field evolves according to
\[
\partial_t \bar B_i + \bar v_k \partial_k \bar B_i - \bar B_k \partial_k \bar v_i =
\partial_k \am_{ik} + \eta \partial_k^2 \bar B_i
\]
where we have defined the antisymmetric matrix 
\[
\am_{ij} \equiv \overline{v'_{[i} B'_{j]}} = \overline{v'_i B'_j} - \overline{v'_j B'_i}
\]
(It is also convenient to define the symmetric matrix
\[
\ap_{ij} \equiv \overline{v'_{\{i} B'_{j\}}} = \overline{v'_i B'_j} + \overline{v'_j B'_i}
\]
although we do not use this here.)
From this we can see that the only sources for a mean field $\bar B$ are a stretching component (which also
appears in the induction equation itself) and a turbulent source term which is more often written as
\[
\partial_k \am_{ik} = \left( \del \times (\overline{v' \times B'}) \right)_i
\]
As we are mainly interested in the zero-mean-field case, we do not discuss the mean-field induction equation
any further.

The turbulent Maxwell-like stress $M_{ij}$ obeys the transport equation
%\bea
%\partial_t M_{ij} + \bar v_k \partial_k M_{ij} - M_{ik} (\partial_k \bar v_j) - (\partial_k \bar v_i ) M_{kj} +
%\partial_k(\overline{B'_i B'_j v'_k} ) =
%\nonumber \\
%- \overline{v'_k B'_j} (\partial_k \bar B_i) - \overline{v'_k B'_i} (\partial_k \bar B_j)
%+ \bar B_k \overline{(\partial_k v'_i ) B'_j} + \bar B_k \overline{(\partial_k v'_j ) B'_i}
%\nonumber \\
%+ \overline{(\partial_k v'_i) B'_j B'_k} + \overline{(\partial_k v'_j) B'_i B'_k}
%+ 
%\eta \left( \partial^2_k M_{ij} - 2 \overline{(\partial_k B'_i)(\partial_k B'_j)} \right)
%\eea
\bea
\partial_t M_{ij} + \bar v_k \partial_k M_{ij} - M_{ik} (\partial_k \bar v_j) - (\partial_k \bar v_i ) M_{kj} +
\partial_k(\overline{B'_i B'_j v'_k}) =
\nonumber \\
-  (\partial_k \bar B^{}_{\{i})\overline{B'_{j\}}v'_k}
+ \bar B_k \overline{(\partial_k v'_{\{i} ) B'_{j\}}}
+ \overline{(\partial_k v'_{\{i}) B'_{j\}} B'_k} +
\nonumber \\
+ \eta \left( \partial^2_k M_{ij} - 2 \overline{(\partial_k B'_i)(\partial_k B'_j)} \right)
\label{eq:max}
\eea

We have so far discussed the Reynolds stress and the turbulent Maxwell stress, but a word is in  order
regarding other approaches to turbulence. In particular we wish to address briefly the dynamics of
turbulent vorticity.
It is tempting to explore the apparent symmetry in the transport of the vorticity $\omega$ and
the magenetic field $B$. In high Reynolds-number flow the vorticity tends to concentrate on
discrete vortex lines. In the extreme case of a superfluid the vorticity of the fluid is identically
zero except on discrete quantized vortex lines, just as in a superconductor the magnetic field
is zero except on discrete quantized field lines; there is certainly an interesting symmetry
to be expolored in the interaction of discrete vortex lines and discrete field lines in the
case of turbulence in a superconducting superfluid. One may suspect that this analogy will
also hold in the case of high Reynolds-number, high magnetic Reynolds-number turbulence.

However, this apparent symmetry is far from perfect. Neither vector
is transported as a passive field. The magnetic field exerts its influence on the bulk flow
through the appearance of the divergence of the Maxwell stress as a bulk force in the 
fluid equations. In contrast, the vorticity does not so much influence the flow as it does
determine it, insofar as the velocity is determined directly from the vorticity through
the Biot-Savart relation, and conversely the vorticity is simply the curl of the velocity.
As a result, a vortex line, if bent, can ``wrap itself up'' by the self-action of its own vorticity, unlike
a field line. 

So, to the extent that vorticity dynamics may be used -- in analogy to polymer dynamics in
solution \citep{Cho:1988} -- as a basis for turbulence models and quasi-viscoelastic behavior in particular, we expect that the
corresponding analogy of turbulent field lines to polymers should be much more fruitful. 
Note also that a stress tensor may be formed directly from the field, whereas the relationship between the
turbulent vorticity and the Reynolds stress is not straightforward. For this reason we do not treat
the dynamics of turbulent vorticity and we resort entirely to simple Reynolds stress modeling for the stress
due to turbulent hydrodynamic motions.

\section{Boussinesq-Like Approximations}
\subsection{Simplest Case: No coupling}
An ordinary differential model, such as the viscoelastic
Maxwell model, contains in principle all that is needed in order to
determine the stress in a fluid, given initial conditions and  a rate-of-strain history. In steady, so-called
 ``curvilineal'' flows with
closed or nearly-closed streamlines (such as the azimuthal Stokes shear flow in the neighborhood of a rotating
sphere), it is sometimes useful to rewrite the solutions of these differential equations in closed form. 

For example, the upper-convected Maxwell model may be written in closed form as
\[
M = {\eta \over \lambda^2} \int_{-\infty}^{t} \left(C_t^{-1}(\tau) - 1\right) \exp\left(-(t-\tau)/\lambda \right)
\,d\tau,
\]
where $C_t^{-1}$ is the Finger tensor
(see \citet{Jose:1990} p.14). Regardless of whether an integral expression may be found for a given model, another closed-form
way of writing the stress for steady, curvilineal flows is to expand the stress as a series solution in the
powers of the (symmetrized) spatial derivatives of the velocity. 
The conventional way of doing this is to make use of powers of the Rivlin-Eriksen tensors $A^{[n]}$, defined below.
%Terms of order $N$ may be written as a sum
%\[
%\sum_{m_1 + m_2 + \dots + m_N = N} C_{m_1 m_2 \dots m_N}(A^{[1]})^{m_1} (A^{[2]})^{m_2} \dots (A^{[N]})^{m_N}
%\]
% where the $C_{\dots}$ are coefficients.
The first Rivlin-Eriksen tensor, $A^{[1]}$, is the symmetrized velocity gradient (defined in terms of the mean velocity $\bar v_i$),
 so, with the addition of a ``pressure''
term proportional to $\delta_{ij}$,  the Boussinesq model may be
written as
\[
\Pi_{ij} = -P_{\rm turb} \delta_{ij} + \mu_{\rm turb} A^{[1]}_{ij},
\]
for incompressible flow. (If the flow is compressible, it is conventional to break the rate-of-strain tensor into
the divergence and a traceless part as we have done ealier; the coefficient of the former is then the turbulent bulk viscosity.)

A Boussinesq-like approximation may be ``derived,'' in a very crude sense, from simple turbulence models 
\footnote{The author is grateful to Michael Steinkamp for pointing this out to him.}. 
A full turbulence model should make consideration of the {\em source} of the turbulence, such as an instability
such as convection or the MRI. However, we point out that a simple turbulent viscosity works well as a first
approximation when solving for the turbulent transport of flow through the lumen of a pipe (i.e. pipe Poiseuille flow), 
for which there is no known instability, linear or otherwise.

The unmodeled Reynolds transport equation for incompressible fluid, under conventional rules of ensemble averaging, 
and ignoring effects such as rotation and thermal convective driving, is
\bea
\partial_t {\Rey}_{ij} + \bar v_k \partial_k {\Rey}_{ij} + (\partial_k \bar v_j) {\Rey}_{ik}
 + (\partial_k \bar v_i){\Rey}_{kj} = \nonumber \\
-\partial_k\overline{(v'_i v'_j v'_k)} - \left(\overline{v'_i \partial_j P'} + \overline{ v'_j \partial_i P'}\right)
 + \nu\left(\partial^2_k {\Rey}_{ij}
- 2\overline{ (\partial_k v'_i)(\partial_k v'_j)}\right).
\eea
(This is simply eqn.~(\ref{eq:rey}) rewritten without the terms in which the magnetic field occurs.)
The left-hand side of this equation represents the advection and distortion of the Reynolds stress by the mean flow.
The right-hand side of the equation is the hard part. The crudest models for the right-hand side of this transport
equation will include terms proportional to $\delta_{ij}$ and terms proportional to ${\Rey}_{ij}$.
Effectively, we have advection and distortion on the left hand side (this distortion is sometimes referred to as a 
source, since it changes the total kinetic energy ${1\over 2} {\Rey}_{kk}$), and on the right hand side we have a
source proportional to $\delta_{ij}$ and a sink proportional to ${\Rey}_{ij}$. 
It is somewhat abusive to call these terms
source and sink terms, but the point is that the  one typically has a positive coefficient and the other typically
has a negative coefficient.
For example, consider the Rotta return-to-isotropy term, which is a model for one of the
effects of the correlations of turbulence with
pressure fluctuations, i.e. the second of the three terms on the right-hand side of the above equation (note that
pressure fluctuations also carry sound away, but this is not included in the Rotta term). This
may be written as, e.g.,
\[
-C_R {\epsilon\over K} \left( \Rey_{ij} - {1 \over 3} \delta_{ij} \Rey_{kk} \right),
\]
where $\epsilon$ and $K$ are the turbulence dissipation rate and the turbulent kinetic energy respectively, and
the Rotta coefficient $C_R$ is a positive constant \footnote{Francis Harlow, personal communication}.
Likewise the viscous part of this transport equation --- i.e. the last term on the right-hand-side --- has a 
diffusive part, which is usually ignored, and a dissipative part that is most simply modeled as being proportional
to the Reynolds stress $\Rey_{ij}$ itself.
%This is true, for example, for the so-called Rotta return-to-isotropy term, as well as any dissipative term which
%is proportional to ${\Rey}_{ij}$ itself.

%For example, consider the model terms
%\[
%{\rm r.h.s} = C_R {\sqrt{K}\over s} \left[{1\over3} \delta_{ij} {\Rey}_{kk} - {\Rey}_{ij} \right] - {\sqrt{K}\over s} {\Rey}_{ij} + \dots
%\]
%where the ellipsis indicates other terms such as a diffusive term and a non-conservative pressure term that we have
%not written.

We thus write
\[
\partial_t {\Rey}_{ij} + \bar v_k \partial_k {\Rey}_{ij} + (\partial_k \bar v_j) {\Rey}_{ik} + (\partial_k \bar v_i){\Rey}_{kj} = 
\aR \delta_{ij} - {\sR^{-1}} {\Rey}_{ij}
\]
where $\aR$ and $\sR^{-1}$ are numerical coefficients (which may depend upon certain characteristics of the flow, such
as the shear, the buoyant convective driving, and so forth).  Note that $\sR$ is a relaxation time.
We may then solve for ${\Rey}_{ij}$, in the case of steady shear (so that the advective operator $\partial_t + \bar v_k \partial_k$ gives zero),
by plugging the above equation back into itself to obtain an infinite series,
\[
{\Rey}_{ij} = \sR \aR \delta_{ij} - \sR^2 \aR A^{[1]}_{ij} + \sR^3 \aR A^{[2]}_{ij} - \sR^4 \aR A^{[3]}_{ij} + \dots
\]
This may be seen from the following relation for $A^{[n]}$:
\[
A^{[n+1]}_{ij} = \partial_t A^{[n]}_{ij} + \bar v_k \partial_k A^{[n]}_{ij} + (\partial_k \bar v_i) A^{[n]}_{kj}
+ (\partial_k \bar v_j) A^{[n]}_{ik}.
\]

In fact, it is not at all necessary to assume that the ``source'' term is proportional to $\delta_{ij}$. If we instead write
\[
\partial_t {\Rey}_{ij} + \bar v_k \partial_k {\Rey}_{ij} + (\partial_k \bar v_j) {\Rey}_{ik} + (\partial_k \bar v_i){\Rey}_{kj} = 
S^{[R]}_{ij} - {\sR^{-1}} {\Rey}_{ij}
\]
where $S^{[R]}_{ij}$ is some unspecified tensor, we obtain
\[
{\Rey}_{ij} = \sR S^{[R]}_{ij} - \sR^2 {\Lop}(S^{[R]})_{ij} + \sR^3 {\Lop}^2(S^{[R]})_{ij}
 - \sR^4 {\Lop}^3(S^{[R]})_{ij} + \dots
\]
where $\Lop$ is the linear operator
\[
{\Lop}(S)_{ij} = (\partial_k \bar v_i) S_{kj} + S^{[R]}_{ik} (\partial_k \bar v_j),
\]
so we may formally write
\[
{\Rey}_{ij} = e^{-\sR \Lop}(\sR S^{[R]})_{ij}.
\]
If the relaxation time $\sR$ is much shorter than the shear time scale (which we denote $1/\gamma_s$), then one expects the series to converge
rapidly.

Despite our formal solution above, let us concentrate on the original case, $S^{[R]}_{ij} = \aR \delta_{ij}$. To first order in the shear one obtains then
\[
{\Rey}_{ij} = \sR \aR \delta_{ij} - \sR^2 \aR (\partial_i \bar v_j + \partial_j \bar v_i)
= P_{\rm turb} - \nu_{\rm turb} (\partial_i \bar v_j + \partial_j \bar v_i).
\]
The exact form of the Reynolds stress in this case, where $\sR \gamma_s \ll 1$, is actually quite sensitive to the
source term, which we have here modeled as being proportional to $\delta_{ij}$. Nevertheless, the approximation that
one obtains --- an effective turbulent pressure and an effective turbulent viscosity --- has proven quite useful over
the past century.

The product $\sR \gamma_s$, which one might call a ``turbulent hydrodynamic Weissenburg-Deborah number,'' is in fact generally
less than or of the order of one in pure hydrodynamic turbulence. One can make the argument in very rough fashion using
eddy-viscosity type arguments. Colliding turbulent blobs lose their identity 
on the order of a coherence timescale, which is on the order of an ``eddy turnover'' timescale, which is of the
order of the turbulent lengthscale $\ell_{\rm turb}$ divided by the turbulent velocity scale
$v_{\rm turb}$. The magnitude of the difference in the corresponding
background mean flow from one side of the eddy to the other is $\ell_{\rm turb} \gamma_s$. If this velocity is greater
than the turbulent velocity, the baackground shear will rapidly disrupt the eddy. Thus $\ell_{\rm turb} \gamma_s < v_{\rm turb}$.
But this implies that $\sR \gamma_s < 1$, if $\sR \sim \ell_{\rm turb} / v_{\rm turb}$.
Likewise, in the case of the continuum approximation in hydrodyanmics, 
the molecular collision frequency must exceed the shear rate for the concept of a Stokes molecular 
viscosity to be terribly useful.

The admittedly very crude argument given above does not hold at all in the case of a high magnetic-Reynolds-number
turbulent magnetic field, however. 
%If
%a picture of turbulence as colliding blobs is inappropriate for hydrodynamic turbulence, it is even more inappropriate
%for MHD turbulence. One of the salient facts about the magnetic field is that, in the case of ideal MHD, integral
%curves of the magnetic field are advected by the fluid. In other words, there is flux-freezing.  
In perfect flux-freezing, a magnetic field line maintains its identity, or ``coherence,'' indefinitely.

In reality, dissipative effects will become important at some point. This is true even in ideal MHD, we expect, if
we extend the word ``dissipative'' to include the effects of turbulent cascades and inverse cascades, which will
take energy on intemediate scales and transport that energy either to very small scales or to larger scales. (Whether energy
that is transported to small scales is ultimately dissipated in the molecular sense is not necessarily relevant.)
It is not at all clear, however, that these quasi-dissipative effects become important on a shear time scale.

Consider the transport equation for the turbulent magnetic cross-correlation tensor $M_{ij}$ as written in
eqn.~(\ref{eq:max}).
%We have
%\bea
%\partial_t M_{ij} + \bar v_k \partial_k M_{ij} - (\partial_k \bar v_i)M_{kj} - (\partial_k \bar v_j)M_{ik} = \nonumber \\
%\cdots + 
%\eta \left( \partial^2_k M_{ij} - 2 \overline{(\partial_k B'_i)(\partial_k B'_j)} \right)
%\eea
%(The ellipsis indicates terms that we have for the moment left out; these terms are written in full
%in the appendix.) 
If we write
\[
\partial_t M_{ij} + \bar v_k \partial_k M_{ij} - (\partial_k \bar v_i)M_{kj} - (\partial_k \bar v_j)M_{ik} =
S^{[M]}_{ij} - (1/\sM) M_{ij}
\]
we obtain, in a similar matter to the above, 
\[
M_{ij} = e^{+\sM \Lop}(\sM S^{[M]})_{ij}.
\]

Again, taking the simplest case, $S^{[M]}_{ij} = \aM \delta_{ij}$, we obtain
\[
M_{ij} = \sM \aM \delta_{ij} + \sM^2 \aM(\partial_i \bar v_j + \partial_j \bar v_i) + \sM^3 \aM A^{[2]}_{ij} + \dots
\]
Note that the signs of the coefficients here do not alternate, so that if the relaxation times were
equal, $\sR = \sM$, then the series solution for $M$ might not be expected to converge as rapidly as that for ${\Rey}$.
In reality, however, we note that for simple shear flow (or ``curvlineal'' flow), both series converge exactly with
only three terms, as $A^{[n]}=0$ for $n>2$.

The full turbulent stress tensor is
\[
\Pi_{ij} = -{\Rey}_{ij} + M_{ij} - \half M_{kk}\delta_{ij},
\]
recalling that the Maxwell stress tensor $\tau^{[Maxwell]}$ (if $|E| \ll |B|$) is
\[
\tau^{[Maxwell]}_{ij} = B_i B_j - {\half} B^2 \delta_{ij},
\]
so that the turbulent Maxwell stress tensor ${\Max}_{ij}$ is
\[
{\Max}_{ij} = M_{ij} - \half M_{kk} \delta_{ij}.
\]
We may then write, again for the simplest case where $S^{[R]}_{ij} = \aR \delta_{ij}$ and $S^{[M]}_{ij} = \aM \delta_{ij}$.
that
\bea
\Pi_{ij} = -\left(\sR \aR + \half \sM \aM + (\gamma_s \sM)^2 \sM \aM\right) \delta_{ij} + \nonumber \\
+ (\sR^2 \aR + \sM^2 \aM) A^{[1]}_{ij}  + (-\sR^3 \aR + \sM^3 \aM) A^{[2]}_{ij}.
\nonumber
\eea
The identification of the term proportional to $\delta_{ij}$ above as a turbulent pressure is not unique, as
the rest of the terms are not all traceless. This ambiguity naturally arises from the identification of
$B^2/2$ as the magnetic pressure. Consider the case $M_{ij} = m \delta_{ij}$. Then the turbulent Maxwell
stress is
\[
{\Max}_{ij} = M_{ij} - {\half}M_{kk}\delta_{ij} = m\delta_{ij} - {3 \over 2} m \delta_{ij} = -\half m \delta_{ij} 
\]
so that the Maxwell stress is proportional to $\delta_{ij}$, which looks like a pressure, except that the
nominal magnetic pressure contribution to the stress is actually $-(3/2)m\delta_{ij}$.
With this caveat in mind, then, we identify the term proportional to $\delta_{ij}$ as a pressure.
This enables us to write
\beq
\Pi_{ij} = -P_{\rm turb} \delta_{ij} + \mu_{\rm turb} A^{[1]}_{ij} + \zeta_{\rm turb} A^{[2]}_{ij},
\label{eq:it}
\eeq
where $\zeta_{\rm turb}$ is the effective turbulent elasticity. 
(One alternative choice for defining the turbulent pressure is to assume that the pressure is proportional
to the trace of the full stress tensor, so that
\[
\Pi_{ij} = -P_{\rm turb} \delta_{ij} + \mu_{\rm turb} A^{[1]}_{ij} + \zeta_{\rm turb} 
\left(A^{[2]}_{ij}-{1\over 3}A^{[2]}_{kk}\delta_{ij}\right);
\]
it is not immediately clear if this is a preferable definition or not.)
We may then define an effective turbulent
relaxation rate as
\[
s_{\rm eff} = {\zeta_{\rm turb} \over \mu_{\rm turb}}
\]
and an effective turbulent Weissenburg number
\[
{\rm We}_{\rm turb} = \gamma_s s_{\rm eff}
\]
The elastic term might also be thought of as a higher-order viscous term, for this restricted case of
steady curvilineal flow.

\subsection{Inclusion of coupling between Reynolds and turbulent Maxwell stresses}
In fact, we also obtain an expression of the form (\ref{eq:it}) if we take the more general case that 
\[
\partial_t {\Rey}_{ij} + \bar v_k \partial_k {\Rey}_{ij} + (\partial_k \bar v_j) {\Rey}_{ik} + (\partial_k \bar v_i){\Rey}_{kj} = 
\aR \delta_{ij} + b_{\rm RM} M_{ij} - {\sR^{-1}} {\Rey}_{ij}
\] \[
\partial_t M_{ij} + \bar v_k \partial_k M_{ij} - (\partial_k \bar v_i)M_{kj} - (\partial_k \bar v_j)M_{ik} =
S^{[M]}_{ij} + b_{\rm MR} {\Rey}_{ij} - (1/\sM) M_{ij}
\]
That is to say, in this case, we make use of the available tensors $\delta_{ij}$, ${\Rey}_{ij}$,
and $M_{ij}$. 
The new terms (with coefficients $\bRM$ and $\bMR$) represent exchange terms. As such,
the coefficients must have opposite signs. 
Such a form may be motivated on the general grounds that energy should be exchanged between the Reynolds
stress and the turbulent Maxwell stress; further motivation is found through a more careful analysis of the
transport equations as discussed below.

Let us rewrite the $M_{ij}$ transport equation, highlighting a few terms, and replacing the resistive term
with a simple dissipative term. We have
\bea
\partial_t M_{ij} + \bar v_k \partial_k M_{ij} - M_{ik} (\partial_k \bar v_j) - (\partial_k \bar v_i ) M_{kj} +
 =
\nonumber \\
- {\grey (\partial_k \bar B^{}_{\{i})\overline{B'_{j\}}v'_k} }
+ {\grey  \bar B_k \overline{(\partial_k v'_{\{i} ) B'_{j\}}} }
-{\green \partial_k(\overline{B'_i B'_j v'_k})}
+ {\green  \overline{(\partial_k v'_{\{i}) B'_{j\}} B'_k} } - \sM^{-1} M_{ij}
\eea

Terms written in light {\grey grey} are  those in which the mean field $\bar B_i$ appears.
The terms in {\green green} exchange energy between $\rey_{ij}$ and $M_{ij}$ through the respective terms
$v'_k \partial_k B'_i$ and  $B'_k \partial_k v'_i$ that appear in the induction equation.
(It might seem that we should discuss energy exchange between the Reynolds stress $\rey$ and the Maxwell stress $\Max$
rather than between $\rey$ and $M$.
However, note that
the turbulent energy density $U_{\rm turb}$ is
\beq
U_{\rm turb} = \half \rey_{kk} - \half \Max_{kk} = \half \rey_{kk} + \half M_{kk},
\eeq
so that we recover that the turbulent field energy density is $(\overline{B'_k B'_k})/2$, as it should be.)

Similarly, for the Reynolds stress, we write
\bea
\partial_t R_{ij} + \bar v_k \partial_k R_{ij} + (\partial_k \bar v_i)R_{kj} + R_{ik}(\partial_k \bar v_j)
+ \bar \rho \partial_k (\overline{ v'_i v'_j v'_k }) = 
\nonumber \\
- \overline{v'_{\{i} \partial^{}_{j\}} P'} + 
{\grey (\partial_k \bar B^{}_{\{i}) \overline{v'_{j\}} B'_k}} + 
{\grey \bar B_k \overline{(\partial_k B'_{\{i} ) v'_{j\}} }} +  
{\green \overline{B'_k (\partial_k B'_{\{i}) v'_{j\}}}} -
\nonumber \\
- {\grey \overline{v'_{\{i}\left(\partial^{}_{j\}}(\bar B_k B'_k)\right)}} - 
{\green \half \overline{v'_{\{i}\left( \partial^{}_{j\}}(B'_k B'_k)\right)} } 
- \sR^{-1} R_{ij}
\eea
Again, $\rey$--$M$ exchange terms are highlighted in {\green green}. 
 
One would expect the exchange terms to be proportional to the mean of $v' \cdot (B' \cdot \grad B' - \half \grad (B')^2)$; this is
easily confirmed. The trace of the (zero mean-field) exchange terms in the Reynolds stress transport equation is
\beq
{\rm Tr}(terms) = 2 \overline{v'_\ell B'_k \partial^{}_k B'_\ell} - \overline{v'_\ell \partial^{}_\ell (B'_k B'_k)}
\eeq
whereas the trace of the exchange terms in the Maxwell-like (i.e. $M$) stress transport equation is
\beq
{\rm Tr}(terms) = 2 \overline{v'_\ell B'_k \partial^{}_k B'_\ell} - \overline{v'_\ell \partial^{}_\ell (B'_k B'_k)} + 
2 \partial_k \left[ \overline{ ( v'_{[\ell} B'_{k]} ) B'_\ell} \right]
\eeq
this last term being in conservative form and so identically zero in the homogeneous case.

In a more sophisticated model, the exchange between $\Rey_{ij}$ and $M_{ij}$ might be modeled through
a transport equation for the triple-correlation exchange terms described above. Here, however, we simply
model this exchange in the transport equations for $\Rey_{ij}$ and $M_{ij}$ by assuming that the exchange
is proportional to these tensors themselves.

With this assumption, ${\Rey}_{ij}$ and $M_{ij}$ have solutions of
the form
\[
{\Rey}_{ij} = r_0 \delta_{ij} + r_1 A^{[1]}_{ij} + r_2 A^{[2]}_{ij}
\]
\[
{M}_{ij} = m_0 \delta_{ij} + m_1 A^{[1]}_{ij} + m_2 A^{[2]}_{ij}
\]
%(the specific solutions for the coefficients $r_{0,1,2}$ and $m_{0,1,2}$ are lengthy and are not reproduced here).
So, the expression (\ref{eq:it}) is again obtained. Specifically,
the solution is 
\[ r_0 (1 - \sR \bRM \sM \bMR) = \sR \aR + \sR \bRM \sM \aM)
\] \[ m_0 (1 - \sR \bRM \sM \bMR) = \sM \aM + \sM \bMR \sR \aR
\] \[ r_{i+1} (1 - \sR \bRM \sM \bMR) = \sR \bRM m_i - r_i
\] \[ m_{i+1} (1 - \sR \bRM \sM \bMR) = - \sM \bMR r_i + m_i
\]

\subsection{Effects of Rotation and Stratification}
It might be expected that rotation is not something to be put into a turbulence transport model; it ought to
exist already in the trasport model, so that rotation is just a matter of doing the proper coordinate transformation. 
This philosophy is also in keeping with the principle of material frame indifference, according to
which a fluid should not know (at least as far as its constitutive relations are concerned) about the overall rotation
of the system. Actually the issue is a bit more subtle than this, since the form of the Reynolds and Maxwell stress
transport equations described above (or, typically, in their unmodeled state) will be invariant with respect to
transformation to a rotating reference frame, but this means that the action of the Coriolis force on the 
turbulent blobs in the rotating reference frame is ignored (see, e.g., the discussions in \citet{Pope:2000}
and in \citet{Rudiger:1989}). Inclusion of the Coriolis force may appear as the introduction of new terms with
symmetrized products of the rotation tensor $\Omega_{ij} \equiv \epsilon_{ijk} \Omega_k$, as described below.
%Inclusion of this Coriolis force before the averaging
%is performed to produce the Reynolds-stress transport equation will produce terms proportional to the symmetrized
%product of the Reynolds stress and the antisymmetric rotation tensor $\Omega_{ij}$.
Such considerations are important for a model of the MRI, as well as for the more simple case of Taylor
instability when the Rayleigh stability criterion is not satisfied.
 %We leave this project for a future paper, but we note
The inclusion of the effects of rotation is an ongoing project that we intend for a future paper. % we note here
These considerations must be addressed in order to reproduce the result from simulations that the $rr$ component
of the stress is larger than the $zz$ component of the stress, not to mention the presence of the MRI itself.

It would also be nice to address the effects of stratification on the turbulence, such as is a nessesary ingredient
for the treatment of convective MHD turbulence. Consideration of turbulent elasticity 
should be important for the turbulence in the 
solar convective zone (SCZ), for example, even though the turbulence in the SCZ is not driven (primarily, at least)
by MHD instabilities but rather purely hydrodynamical ones. For incompressible but variable-density stratified fluids
in the laboratory, one finds an additional driving term in the Reynolds stress transport equation that is proportional
to the product of the pressure gradient and the density gradient. In cases of astrophysical interest, where compressiblity
becomes important, the important consideration is the direction of the entropy gradient (or, the gradient in the
potential temperature). This will necessarily introduce anisotropies into the turbulent stress tensor, just as
a consideration of the sources of the MRI will produce anisotropies that are not adequately addressed by the considerations
outlined in this paper, as mentioned above. Inclusion of these effects may appear as the introduction of new terms
with symmetrized products of the vectors $g_i$ and $\p_i \bar P$ and $\p_i \bar \rho$, or other thermodynamic derivatives,
depending upon the choice made by the modeler for the independent thermodynamic quantities.
Again, a proper inclusion of these effects must follow from a model
that is built from the transport equations in which --- unlike what we have presented here --- the assumption of
incompressibility is relaxed. This is also an ongoing project that we intend for a future paper.

We can, however, comment briefly upon the combined effects of rotation and stratification as they manifest
themselves in stellar convective zones through the so-caled
``$\Lambda$-effect'' \citep{Rudiger:1989}. This introduces additional terms that, in the SCZ, are to lowest order proportional
to the dyad product $(\vec g \times \vec \Omega)\vec g$. 
%(We are not aware of how best to generalize this to cases
%where, unlike the  typical case in
%stellar convective zones,
% the gravity vector $\vec g$ is not nearly collinear with $\grad P$ and $\grad \rho$, and so for what follows,
%we will be assuming that we are discussing convection within the context of a stellar convective zone.
(Note that inclusion of the rotation $\Omega_i$ violates one of the assumptions 
that forms the basis of the expansion in terms of
Rivlin-Eriksen tensors, namely material frame indifference, but again, this is to be expected for a turbulence
model that takes account of Coriolis forces, as mentioned above.)

For the $\Lambda$-effect as described by \citet{Rudiger:1989} (in particular, section 4.6), 
the additional terms appear in the prescription
for the equilibrium Reynolds stress (as opposed to the Reynolds stress transport equation), and for the
lowest-order $\Lambda$-effect the symmetrized product mentioned above,
\[
\rey^{\rm lambda}_{ij} = \bar \rho \Lambda_O \left(g_i g_k \Omega_{kj} + \Omega_{ik}g_k g_j\right) + \dots,
\]
produces additional components $\rey^{\rm lambda}_{r\phi}$ and, to higher order,
 $\rey^{\rm lambda}_{\theta \phi}$. These decouple from the
elastic terms in the stress, so that the total turbulent stress tensor may be written as
\[
\Pi_{ij} = -P^{\rm turb}_{ij} + \Pi^{\rm visc}_{ij} + \Pi^{\rm elastic}_{ij} + \Pi^{\rm lambda}_{ij}.
\]
The form of the turbulent stress tensor may then be written as an expression of the form
(compare \citet{RekRue:1998} eqn.~17)
\beq
\Pi_{ij} = -P \delta_{ij} + \bar \rho {N}_{ijk\ell}(\p_k\bar v_\ell) +
 \bar \rho Z_{ijk\ell mn} (\p_k \bar v_\ell)(\p_m \bar v_n) - \bar \rho \Lambda_{ijk}\Omega_k
\eeq
so that the various effects are simply additive. 

\section{Conclusions}
We have discussed some modifications to the basic Boussinesq ansatz that are motivated by an analogy between
the turbulent component of the magnetic field in MHD turbulence and the dynamics of polymers in solution. This
is part of an ongoing project begun by us to study MHD turbulence beginning with the Reynolds decomposition
of the field, with particular focus on the small-scale tangled component in preference to the large-scale mean-field component.
It is our belief that undue focus on the latter, apparently for the sake of understanding dynamo behavior, has obscured
the important dynamical effects that the former may have in practical MHD turbulent flow problems, and that furthermore
the Boussinesq effective viscosity had been adopted for such flows without sufficient consideration as to whether the
initial motivation for an effective turbulent viscosity carried over from hydrodynamics to MHD.
While a viscoelastic picture of MHD turbulence, built from --- or at least inspired by --- a polymer analogy, may not
be ``correct'' in some absolute sense (as pointed out by a previous anonymous referee), we feel that it is at
least {\em useful}, which is all that a model can ask to be.
Of course, there are many effects that naturally arise from viscoelastic models, as described at
length in any textbook on the matter, which we hope to discuss in future work.
 For example, the turbulent medium supports Alfv\'en waves that can be treated
easily within a viscoelastic picture; these turbulent Alfv\'en waves have a complex dispersion relation, reflecting the
dissipative nature of turbulence. These waves do not appear at all in a purely viscous model. We take this as a further
indication that a viscoelastic picture of MHD turbulence may be useful.

Here we have focused on one aspect common to any viscoelastic model of MHD turbulence, which is to say any model that takes
into account the distortion of the turbulent Maxwell stress by the inclusion of the proper tensor derivatives and a
relaxation time, namely the appearance of significant anisotropy of the turbulent stress tensor in the direction
of the shear. This effect should be important in any shearing environment in MHD turbulence, including but not
limited to the MRI (and in particular the inner regions of accretion disks and the jet-launching process) and to
the turbulence in stellar convection zones.

\end{document}